\documentclass[authoryear,preprint]{elsarticle}

\usepackage{natbib}
\usepackage{lineno}
\usepackage{hyperref}
\usepackage{amssymb}
 \usepackage{graphicx}
 \usepackage{amsmath}
 \usepackage{xcolor}
 \usepackage{multirow}
\nolinenumbers

\journal{Expert Systems with Applications}

\newpageafter{author}

\begin{document}
\nolinenumbers

\begin{frontmatter}

\title{Ordinal classification for interval-valued data and interval-valued
functional data}


\author[mymainaddress]{Aleix Alcacer}
\ead{aalcacer@uji.es}

\author[mymainaddress]{Marina Martinez-Garcia}
\ead{martigar@uji.es}

\author[mymainaddress]{Irene Epifanio\corref{mycorrespondingauthor}}
\cortext[mycorrespondingauthor]{Corresponding author}
\ead{epifanio@uji.es}

\address[mymainaddress]{Dep. Matem\`atiques, Universitat Jaume I, Castelló 12071, Spain}

\begin{abstract}
The aim of ordinal classification is to predict the ordered labels of the output from a set of observed inputs. Interval-valued data refers to data in the form of intervals.  For the first time, interval-valued data and interval-valued functional data are considered as inputs in an ordinal classification problem. Six ordinal classifiers for interval data and interval-valued functional data are proposed. Three of them are parametric, one of them is based on ordinal binary decompositions and the other two are based on ordered logistic regression. The other three methods are based on the use of distances between interval data and kernels on interval data. One of the methods uses the weighted $k$-nearest-neighbor technique for ordinal classification. Another method considers kernel principal component analysis plus an ordinal classifier. And the sixth method, which is the method that performs best, uses a kernel-induced ordinal random forest. They are compared with na\"ive approaches in an extensive experimental study with synthetic and original real data sets, about human global development, and weather data. The results show that considering ordering and interval-valued information improves the accuracy. The source code and data sets are available at \url{https://github.com/aleixalcacer/OCFIVD}.
\end{abstract}


\end{frontmatter}

\nolinenumbers

\section{Introduction}
Symbolic data analysis (SDA) is gaining popularity since this kind of data can arise as the result of aggregation of very large data sets, 
which are very common nowadays  \citep{doi:10.1198/016214503000242}. However, there are also data that are naturally symbolic (see \citet{billard2008some} for some illustrative examples of symbolic data). In classical multivariate analysis, data points consist of a single (numerical or categorical) value for each feature, i.e. they are point-valued data. 
However, symbolic data can be lists, intervals,
histograms, etc.

In this work, we focus on interval-valued data (IVD), i.e. each data point is expressed in interval format. Using classical techniques with IVD can cause distorted results due to the loss of information, as explained in \citet{billard2008some}. \textcolor{black}{Each single observation of point-valued data has  no internal variation unlike any single symbolic data value, which has its own internal variation \citep{billard2006symbolic}. For example, for an interval-valued observation $[4, 10]$ and assuming a uniform distribution across the interval, the variance ($s^2$) is 3 (see \citet{bertrand2000descriptive} on how to calculate the symbolic sample variance). However, if the mid-point of that interval (7) is considered as a point-valued data, $s^2$ = 0.}

\textcolor{black}{As consequence, the sample variance of the entire interval-valued data set
comprises both the within- and between- observation variations. As an illustrative example, \citet{billard2006symbolic} compared the results obtained by using symbolic analysis versus classical analysis with principal component analysis (PCA). The classical results  were less informative than 
the richer knowledge gained from the symbolic analysis. However, this is not exclusive to  PCA, rather the use of naive ways to deal with IVD may introduce errors in analysis  \citep{li2019matrix}.} 
This is why methodologies for IVD should be used with this kind of data. 

\subsection{\textcolor{black}{Interval-valued data}}
\textcolor{black}{IVD data have been used in different statistical problems, such as archetypal analysis \citep{d2012interval}, classification \citep{silva2006linear,silva2015discriminant, RAMOSGUAJARDO2016591, appice2006classification,  de2008multi,  JAHANSHAHLOO2007521, 8963733, ANGULO20081220, rossi2002multi}, clustering \citep{10.1007/978-3-642-22194-1_76,SUN2022910, chen2019study, d2023wavelet}, hypothesis testing \citep{grzegorzewski2019sign, maharaj2021test}, outlier analysis \cite{duarte2018outlier}, PCA \citep{lauro2000principal, le2012symbolic,  SUN202294}, and regression analysis \citep{blanco2011estimation,sinova2012interval,xu2022bivariate}. An excellent survey is found in \citet{brito2014symbolic}. }

Recent advances in analysis of IVD  and interesting applications comprise a variety of fields such as carbon price forecasting \citep{LIU2022116267}, $PM_{2.5}$ concentration forecasting \citep{WANG2022117707}, the spatial behavior of the number of cases per COVID-19 and rent price analysis \citep{FREITAS2022116561}, clustering Fungi species \citep{RIZORODRIGUEZ2022116774}  and forecasting of oil price \citep{SUN2022772}. 

\textcolor{black}{Focusing on classification methodologies,   \citet{silva2006linear}, \citet{silva2015discriminant} and \citet{RAMOSGUAJARDO2016591} developed discriminant analysis methods for interval data. \citet{appice2006classification} considered different distances
 for several types of symbolic data with the $k$ nearest neighbor method. \citet{de2008multi} introduced multi-class logistic regression models. \citet{JAHANSHAHLOO2007521} extended data envelopment analysis–discriminant
analysis methodology for interval data. \citet{8963733} applied traditional classification methods with a different representation of interval data.  Support vector machines \citep{ANGULO20081220}  and artificial neural networks \citep{rossi2002multi} were used in interval data classification.}

Another field that is gaining popularity is functional data analysis (FDA), since technological advances have permitted the acquisition of functional data. In FDA, data points are functions. \citet{Ramsay05}  provide an excellent overview of FDA. 
Usually, the observed functions are point-valued, but they can also be interval-valued functions (IVF). Some examples of these are the maximum and minimum 
temperatures for each day in meteorological stations, daily interval stock prices,  and a  person's daily blood pressure records. \textcolor{black}{Working with the intervals provides a more realistic view on the weather conditions variations than the simple average values. Analogously, the intervals offer a more relevant information for experts in order to evaluate the stocks tendency and volatility in the same day \citep{lauro2000principal}. Similarly, being aware of the blood pressure fluctuations is critical from the healthy point of view \citep{sanidas2019labile}. Likewise, more meaningful results are obtained when interval-valued functions are used rather than single-point functional data, as shown  by  \citep{d2023wavelet} in clustering.}

\subsection{\textcolor{black}{Ordinal classification} \label{taxonomia}}
In classification, a label in a set has to be predicted based on the observation of several inputs. Most of the time, labels are considered unordered, even if they are not. Therefore, the vast majority of classification algorithms are conceived to solve nominal classification problems. Nevertheless, ordered classification problems arise in many fields, such as collaborative filtering \citep{alcacer2021combining}, environmental management \citep{balugani2021dimensionality},  finance \citep{hirk2019multivariate}, information retrieval, 
medicine \citep{barbero2021ordinal,singer2021classification}, psychology, and social sciences, among others \cite{survey}. In ordered classification problems, labels are ordered: for example, ratings (very low, low, indifferent, high, very high). \textcolor{black}{In those problems, misclassifying an instance in a neighboring class is generally less relevant than misclassifying it in distant classes. Furthermore, \citet{vargas2022unimodal} explained that taking order into account in ordinal classifiers usually accelerates the learning process and reduces the amount of data needed for training. }

\textcolor{black}{A taxonomy of ordinal classification methods was proposed by \citet{survey}, where methodologies were grouped into three approaches. Na\"ive approaches use simpler paradigms and comprise regression, nominal classification (the order is neglected \citep{ferrando2020detecting}), and cost-sensitive classification. The second approach divides the ordinal categories into several binary labels, i.e. it uses binary decompositions, as made by \citet{Frank_2001}. Finally, the third approach comprises the threshold models, which are the most successful methods in ordinal classification. They assume that there is a continuous feature that can explain the behavior of the ordinal factor. These models include:  cumulative link models \citep[Ch. 7]{agresti2002categorical}, support vector machines, discriminant learning, perceptron learning, augmented binary classification, ensembles \citep{hechenbichler2004weighted,ordinalRF,vega2021ocean}, and Gaussian processes.  }

The number of methods for ordered classification for multivariate data is much lower than for nominal classification \citep{ordinalRF,pierola2016ensemble}, and even extremely lower for functional data \citep{FERRANDO_2021} or other kinds of data, such as those on Riemannian manifolds \citep{Simo2020}. For the case of interval data, 
literature about ordinal classification is very scarce. In fact, to the best of our knowledge, only monotonic classification, which is a particular case of ordinal classification, has been studied for IVD \citep{chen2022hybrid}.  In monotonic classification, there are monotonicity constraints between inputs and outputs. 

\subsection{\textcolor{black}{Our contributions}}
Due to this scarcity, the objectives of this work are to introduce and compare different ordinal methods for IVD and IVF. We also aim to compare them with the na\"ive approach, which consists of a) not taking the order into account, i.e. applying nominal classification methods to ordinal classification problems as if classes were unordered, or b) taking order into account, but discarding the interval-valued nature of the data. \textcolor{black}{\citet{survey} carried out a comparative evaluation of ordinal methods for multivariate data and showed that, even if the na\"ive approach can be very competitive, taking into account order improves the performance.} To the best of our knowledge, no previous work has had these objectives. Therefore, the novelty of this work lies in:

\begin{itemize}
    \item Proposing several methods for ordinal classification of IVD and IVF.
     \item Comparing those methods using simulated and real data.
        \item  \textcolor{black}{Comparing} the proposed ordinal methods using nominal classification methods for IVD and IVF to see whether not discarding the information about order is important for the case of IVD. 
         \item  \textcolor{black}{Comparing the performance when using ordinal classifiers, but without taking into account the interval-valued information of the data, i.e. considering the mid-point of the intervals as inputs.  }
       \item Providing the R \citep{R} code with the algorithms. The data and code are available at \url{https://github.com/aleixalcacer/OCFIVD} for reproducibility. 
     \end{itemize}

The paper is  organized as follows: Section \ref{current} reviews previous works that will be used in our proposals. Section \ref{metodologia} introduces the proposed ordinal  classification methods for both IVD and IVF. The results are presented and discussed in Section \ref{resultados}. Finally, Section \ref{conclusiones} contains some conclusions.


\section{Current methodologies \label{current}}
\subsection{Multivariate ordinal classification methodologies} We review ordinal classifiers for point-valued data. Let $\bf{X}$ be an  $N \times K$ matrix with  $N$ observations (${\bf x}_i$) of $K$ features and $\bf{y}$ the response vector, which is an ordered factor with $Q$ levels, the ordered classes ${C_1, ..., C_Q}$.

\begin{description}
\item [The Frank and Hall (FH) method \citep{Frank_2001}:] the ordinal classification problem is broken into a series of $Q-1$ binary classification problems, where one class is made up of ${C_1, ..., {C_q}}$ and the other class of ${C_{q+1}, ..., C_Q}$, for ${q}$ = 1 ,..., $Q-1$. For a new observation with features $\bf{x}$, let $p_q$ be the estimate of $P( y > C_q | {\bf x})$. Therefore, the predicted probabilities  of each of the $Q$ classes are: 
 $P( y = C_1 | {\bf x})$ = $1 - p_1$; $P( y = C_q | {\bf x})$ = $p_{q-1} - p_q$, $q$ = 2, ..., $Q - 1$;  $P( y = C_Q | {\bf x})$ = $p_{Q-1}$, for $q$ = 1, ..., $Q - 1$. In order to apply this method, we need the binary classifier to return  class probability
estimates. 
\item [Weighted $k$-Nearest-Neighbor Techniques for Ordinal Classification 
($wk$NN):] \citet{hechenbichler2004weighted} describe the use of weighted nearest neighbors for ordinal classification. For a new observation $\bf{x}$, the $k+1$ nearest neighbors to $\bf{x}$ are found according to a certain distance function $d({\bf{x}}, {\bf{x}}_i)$. The $k$ smallest distances are normalized by dividing them by the distance to the ($k$ + 1)th neighbor and transformed into weights by any kernel function. In the case of nominal classification, the predicted class is usually chosen by the weighted majority vote of the $k$ nearest neighbors, i.e. by the mode. But for ordinal classification, the weighted median is used for predicting the class. 
\textcolor{black}{The implementation is based on the function $kknn$ from the R package {\bf kknn} \citep{kknn}.}



\item [Ordered logistic regression (POLR):]  
The cumulative link model is described in detail in  \citet[Ch. 7]{agresti2002categorical}. The model is $logit$ $P( y <= C_q | \bf{x})$ = $\zeta_q$ - $\eta$, where the inverse of the standard logistic cumulative distribution function is the logit link function ($logit(p)$ = $log(p/(1-p))$), 
 $\zeta_q$ parameters give each cumulative logit, and $\eta$ represents the linear predictor $\beta_1 x_1$ + ... + $\beta_K x_K$. 
 After estimating the parameters, we can predict the class probabilities for a new instance, which is assigned to the class with the highest probability. We have
used the  $polr$ 
function from the R package {\bf MASS} \citep{mass} in the implementation.

\item [Ordinal forest (OF):] In OF, ordinal classes are predicted by a  random forest methodology introduced by \citet{ordinalRF} for
 multivariate features. Optimized score values are used in place of the category values of the ordinal response and the results are treated as a metric output. The method is implemented in the function {\it ordfor} of the R package {\bf ordinalForest}.

\end{description}

\subsection{Point-valued functional ordinal classification methodologies \label{point}}
\begin{description}
\item [Kernel-Induced Random Forests (KIRF):] KIRF was introduced by
\cite{DBLP:conf/cibcb/FanCW10} for nominal classification of functional data.
Rather than using the raw observations, the idea is to use kernel functions of each two different observations of the training set as candidate splitting rules in the kernel-induced classification trees. Those trees are employed in KIRF.

\item [Functional principal component analysis (FPCA) + ordinal classifier:] \cite{aguilera2008solving} considered FPCA followed by POLR. In \cite{FERRANDO_2021} more methodologies for functional ordinal classification based on a FPCA decomposition of the functions followed by an ordinal classifier are considered. 
\end{description}

\subsection{Interval-valued classification} 
We focus on \textcolor{black}{the methodologies related with our proposals}. \textcolor{black}{\citet{de2008multi} proposed  applying multi-class logistic regression model in two ways. First, they fit that model jointly to the lower and upper extreme values of
the intervals. Second, they fit the model to the lower and upper
extreme values of
the intervals separately. }

The methodology proposed by \citet{silva2015discriminant} returns class probabilities, which are needed for the FH method. In that method, intervals are represented by a bivariate normal distribution (one feature is for the midpoints and the second feature for the logarithm of ranges of the intervals), and classical linear (or quadratic) discriminant analysis are applied. 
  We refer to this method as LDA-ID (Linear Discriminant Analysis of Interval Data). 

\textcolor{black}{ Let $\bf{X}$ be an  $N \times K$ matrix with  $N$ observations (${X}_i$) of $K$ interval variables (${X}_{.j}$), i.e. $X_{ij}$ = $[x_{ij}^l,x_{ij}^u]$. Assuming all intervals are non-degenerate ($x_{ij}^l < x_{ij}^u$, $i$ = 1, ..., $N$; $j$ = 1, ..., $K$),  $X_{ij}$ is represented by the midpoint $c_{ij}$ = $(x_{ij}^l + x_{ij}^u)/2$ and the log-range $r_{ij}^*$ = $ln(x_{ij}^u - x_{ij}^l)$.   In the Gaussian model, a joint multivariate Normal distribution N($\bf{\mu},\bf{\Sigma}$) is assumed  for the
midpoints $\bf{C}$ and the logs of the ranges $\bf{R^*}$ i.e.,  $\bf{\mu}$ = $[{\bf{\mu}}_C' \hspace{.2cm}  {\bf{\mu}}_{R^*}']$ and $\bf{\Sigma}$ = $\left( 
\begin{matrix} {\bf{\Sigma}}_{CC} & {\bf{\Sigma}}_{CR^*} \\ {\bf{\Sigma}}_{R^*C} &{\bf{\Sigma}}_{R^*R^*} \end{matrix}\right)$,   where $'$ denotes the transpose and  ${\bf{\mu}}_C$ and ${\bf{\mu}}_{R^*}$ are $K$-dimensional column vectors of the mean values of $\bf{C}$ and $\bf{R^*}$, respectively, and ${\bf{\Sigma}}_{CC} $, ${\bf{\Sigma}}_{CR^*}$, ${\bf{\Sigma}}_{R^*C}$  and ${\bf{\Sigma}}_{R^*R^*}$ are their variance-covariance matrices. Four different configurations for the variance-covariance matrix are considered: case 1) non-zero correlations among
all $\bf{C}$ and $\bf{R^*}$, therefore, there is no restriction on ${\bf{\Sigma}}$; case 2) variables ${X}_{.j}$ are not correlated, therefore, ${\bf{\Sigma}}_{CC} $, ${\bf{\Sigma}}_{CR^*}$, ${\bf{\Sigma}}_{R^*C}$  and ${\bf{\Sigma}}_{R^*R^*}$ are diagonal; case 3) there is no correlation between $\bf{C}$ and $\bf{R^*}$, therefore, ${\bf{\Sigma}}_{CR^*}$ = ${\bf{\Sigma}}_{R^*C}$ = 0; case 4) all $\bf{C}$ and $\bf{R^*}$ are not correlated, therefore,  $\bf{\Sigma}$ is diagonal. See \citet{brito2012modelling} for details about the estimators for each covariance configuration. For each configuration, the classical linear  discriminant classification rule can be obtained as explained by \citet{silva2015discriminant}.}

\textcolor{black}{The methodology is implemented in the function {\it lda} of the R package  {\bf MAINT.Data} \citep{RJ-2021-074}.}  \textcolor{black}{Results for all four configurations are compared by the Bayesian
Information Criterion (BIC), and the one with the lowest BIC value is selected. } 

\subsection{Distances between interval-valued data and interval-valued functions \label{distancias}}
Several distances have been defined for IVD and IVF \citep{10.1007/978-3-642-22194-1_76,  SUN2022910, SUN202294}. The Hausdorff distance is commonly used. Let us review its definition.

Let $X_i$ = ($[x_{i1}^l,x_{i1}^u]$, ..., $[x_{iK}^l,x_{iK}^u]$) be an observation of $K$-dimensional IVD, where $i$ = 1, ..., $N$ and $x_{ik}^l \leq x_{ik}^u$, $x_{ik}^l$, $x_{ik}^u$ $\in$ $\mathbb{R}$ . The Hausdorff distance between $X_i$ and $X_j$ is defined by $D^H(X_i,X_j)$ = $\sum_{k=1}^K D_k(X_i, X_j)$, where $D_k(X_i, X_j)$ = $max(|x_{ik}^l - x_{jk}^l| , |x_{ik}^u - x_{jk}^u| )$. The Euclidean Hausdorff Distance is defined by $D^{EH}(X_i,X_j)$ = $\sqrt{\sum_{k=1}^K [D_k(X_i, X_j)]^2}$. 

Analogously, the functional Hausdorff distance is defined for a set of IVF $X_i (t)$ = $[x_i^l (t), x_i^u (t)]$, with $i$ = 1, ..., $N$, $x_i^l (t) \leq x_i^u (t)$ and $t$ $\in$ $[a, b]$, i.e. $x_i^l (t)$ ($x_i^u (t)$) is the lower (upper) function  of $X_i (t)$. The functional Hausdorff distance between $X_i (t)$ and $X_j (t)$ is defined by $D^{FH}(X_i (t) ,X_j (t) )$ = $\int_a^b max(|x_{i}^l (t) - x_{j}^l (t)| , |x_{i}^u (t) - x_{j}^u (t)| ) dt$. The Functional Euclidean Hausdorff distance is defined by 
$D^{FEH}(X_i (t) ,X_j (t) )$ = $\int_a^b \sqrt{max\{|x_{i}^l (t) - x_{j}^l (t)|^2 , |x_{i}^u (t) - x_{j}^u (t)|^2 \}} dt$. In practice, the integral can be estimated by numerical integration such as the trapezoidal rule. If we have multivariate IVF, for example, bivariate IVF $F_i (t)$ = $(X_i (t), Y_i (t))$  we can define $D^{FEH}(F_i (t),F_j (t))$ =  $\sqrt{ D^{FEH}(X_i(t), X_j(t))^2+ D^{FEH}(Y_i(t), Y_j(t))^2}$. 

\textcolor{black}{If the scale of the variables is very different, some standardization should be carried out. See \citet{de2006dynamic} for different alternatives for IVD.}

\subsection{Kernel on interval data}

\cite{do2005kernel} defined a Radial Basis Function  (RBF) kernel for dealing with interval data as follows: $K_I \langle X_i, X_j \rangle$ = $exp (- D^{EH}(X_i,X_j)^2 / \gamma$), where the parameter $\gamma$ is the spread of the kernel. 

Once the kernel is defined, we can use kernel techniques, such as Kernel Principal Component Analysis (KPCA) \citep{scholkopf1998nonlinear}. \textcolor{black}{The implementation is based on the function $kpca$ from the R package {\bf kernlab} \citep{karatzoglou2004kernlab}.}


\section{Proposed methods for ordinal classification of interval-valued data and interval-valued functions \label{metodologia}}
Besides na\"ive approaches, such as 1) working with midpoints of intervals and applying ordinal classification methods or 2) using interval classification methods without considering order, we propose the following methods that consider both the order in the response and the interval nature of the input data.

\begin{description}
\item [FH+ LDA-ID:] we  propose to use the FH method with LDA-ID as the binary classifier.  Although LDA-ID is intended for IVD, we can also use it with IVF. The idea is to discretize the observed IVF to a (fine) grid
of equally spaced values, so we can work with them as IVD. Therefore, FH + LDA-ID can be used  with both IVD and IVF. Note that FH takes the order of the output into account and LDA-ID obtains the class probabilities for each  interval-valued binary classification problem. If the grid where functions are discretized is too fine, this methodology will not work with functions if we have a great quantity of observed features with respect to observations. In that case, the grid should not be so fine. 


\item [DI+$wk$NN:] our proposal is to use the distances introduced in Sect. \ref{distancias} together with $wk$NN for ordinal classification. Depending on whether we work with IVD or IVF, the method will be  $D^{EH}$+$wk$NN or $D^{FEH}$+$wk$NN, respectively, although they will be referred to as DI+$wk$NN indistinctly. In the implementation, $k$ = 7 is used; the parameter has not been tuned.


\item [KPCA + ordinal classifiers:] our idea is to carry out a feature extraction stage followed by an ordinal classifier for multivariate data. Preprocessing is a powerful methodology in multivariate data \citep[p.150-151]{hastie2009elements} for improving the performance of a learning procedure. Previously, it has also been  employed successfully with functional data \citep{Epitechnometrics,Epifanio11}. The idea is similar to carrying out FPCA + ordinal classifier, which is explained in Sect. \ref{point} for point-valued functional data, but in our proposal we consider KPCA with a kernel for interval data.

Note that 
\citet{do2005kernel} only defined an RBF kernel for IVD. However, we can extend it for IVF. Therefore, we introduce a new definition, an RBF kernel for IVF  $K_{FI} \langle X_i (t), X_j (t) \rangle$ = $\exp{ (- D^{FEH}(X_i (t),X_j (t))^2 / \gamma}$). In the implementation, $\gamma$ = 1 is used; the parameter has not been tuned.

Therefore, depending on whether we work with IVD or IVF, the method will be $K_I$PCA+ ordinal classifier or $K_{FI}$PCA+ ordinal classifier, respectively.  The ordinal classifier can be POLR, OF, or another ordinal classifier for multivariate data. We have considered POLR in the experimental section, so we will refer to KPCA +POLR  for IVD or IVF indistinctly. 


\item[KIOF:] our proposal is to use KIRF with IVD and IVF using $K_I$ or $K_{FI}$, respectively, but rather than using nominal classification of random forest (RF) as in KIRF, we consider OF. We refer to this method   as KIOF for both $K_I$IOF and $K_{FI}$IOF indistinctly. 


\item[POLR-I and POLR-I2:] our proposal is to extend the ideas in \cite{de2008multi} to ordinal classification of IVD and IVF. Rather than using a multi-class logistic regression model, POLR is considered. For IVF, they can be discretized as explained in FH + LDA-ID. 

Therefore, in POLR-I, the inputs of POLR are the lower and upper bounds of the intervals. However, in POLR-I2, we build two models. One model is built by applying POLR to  the lower  bounds of the intervals, while the other model is built by applying POLR to  the upper  bounds of the intervals. The posterior probabilities
of the classes for both models are averaged to obtain the final posterior probabilities
of the classes.

\end{description}

\textcolor{black}{FH+ LDA-ID belongs to the second approach of ordinal binary decompositions, while the rest of methods belong to the third approach of threshold methods. DI+$wk$NN and KIOF are included in the ensembles subgroup, while KPCA +POLR, POLR-I and POLR-I2 are part of the cumulative link models. In the multivariate case, methods of the third approach are the most successful \citep{vega2021ocean}.  Although POLR is fast to train and one of the most popular ordinal classification methods, it has not the best performance in the multivariate setting \citep{survey}. Therefore, a priori, we expect the last five methods proposed being the best performing, and, specially DI+$wk$NN and KIOF,  if IVD (or IVF) case follows the same pattern as the multivariate case \citep{survey}. }

\textcolor{black}{ Table \ref{overview} presents an overview of the methods used in the experiments for IVD (analogously for IVF). On the one hand,} \textcolor{black}{the proposed methodologies (FH+LDA-ID, DI+ $wk$NN, KPCA + POLR, KIOF, POLR-I, and POLR-I2)}. \textcolor{black}{On the other hand, } \textcolor{black}{the established techniques, which are na\"ive methods in this case, since no previous methodology has been considered for ordinal classification of IVD until now. Three na\"ive methodologies are contemplated: using LDA-ID, which is an interval-valued classifier that does not take order into account; and using POLR and OF considering the midpoints of intervals as inputs, i.e. these two classifiers take into account order, but not the interval-valued nature of the data. } \textcolor{black}{In the implementation,   default parameters (they can be seen in the code file) have been considered.} 

\begin{table}[ht]
\caption{\label{overview} \textcolor{black}{Summary of some characteristics of the methods.}}
\centering
\renewcommand{\arraystretch}{1.5}
\begin{tiny} 
\begin{tabular}{p{2.5cm}p{1.5cm}cp{4.5cm}}
Method & Approach \textcolor{black}{(Sect. \ref{taxonomia})} & Parametric&  Input   \\\hline
POLR \textcolor{black}{(Ordered logistic regression)} & 1st & Yes & Midpoints of IVD\\
OF \textcolor{black}{(Ordinal forest)} & 1st & No & Midpoints of IVD \\
LDA-ID  \textcolor{black}{(Linear Discriminant Analysis of Interval Data)}&  1st &  Yes & Interval bounds transformed into midpoints and log-ranges \\
FH+LDA-ID \textcolor{black}{(Frank and Hall method + LDA-ID)} & 2nd & Yes & Midpoints and log-ranges for each binary classification problem of FH-method\\
DI+ $wk$NN \textcolor{black}{(Weighted $k$-Nearest-Neighbor with $D^{EH}$ or $D^{FEH}$ for IVD and IVF resp.)} & 3rd & No & Interval bounds for computing the distances (input of $wk$NN)\\
KPCA + POLR \textcolor{black}{(Kernel Principal Component Analysis + POLR)} & 3rd & Mixed & Interval bounds for computing the distances and the kernel. The projections on the principal components (input of POLR)\\
KIOF \textcolor{black}{(Kernel-Induced Random Forests with OF)} &  3rd & No & Interval bounds for computing the distances and the kernel (input of OF)\\
POLR-I \textcolor{black}{(1st Extension of \cite{de2008multi} with POLR)} & 3rd & Yes & Lower and upper
bounds separately, as two distinct features\\
POLR-I2 \textcolor{black}{(2nd Extension of \cite{de2008multi} with POLR)} & 3rd & Yes & Lower bounds to POLR-1 and upper bounds to POLR-2. Posterior probabilities of POLR-1 and POLR-2 to combine them
\\\hline
\end{tabular}
\end{tiny} 
\end{table}

\section{Results and discussion \label{resultados}}

 We compare the methodologies in three different scenarios. Artificial data are generated in Sect. \ref{simu}, while real data sets are considered in Sect. \ref{deve} and \ref{meteo}, which deal with IVD and IVF, respectively. 

Although the details will be provided for each scenario, the experimental setup is common to all  three scenarios. In each scenario, we consider a variety of ordered levels for the output. A total of 50 full data sets are built in each case. Each of these data sets is divided into a training set and a test set. For assessing performance, we compute accuracy (success rate) in the corresponding test set.  As the same data sets are used to compare all the methods, a completely randomized block design is applied to test the differences between methods, together with Tukey’s test for comparing all pairs of means  \citep{montgomery2019design}. We have used the R package multcomp \citep{Hothorn}.

\subsection{Simulated data} \label{simu}
Two ordinal classification problems with IVD are simulated, with three and four ordered classes, respectively. The simulation design of synthetic interval data sets resembles that  made by \citet{de2008multi}. Nevertheless,  we consider more than 3 classes and the distribution parameters are different bearing in mind that the output is ordinal. 

Each class is composed of 100 samples generated according to bivariate Normal distributions with non-correlated components and the following parameters for the mean ${\bf {\mu}}$ and the covariance matrix  ${\bf \Sigma}$: 
${\bf \mu} = \begin{bmatrix}
    \mu_1 \\
    \mu_2
  \end{bmatrix}$
and ${\bf \Sigma} = \begin{bmatrix}
    \sigma^2_1 & \rho \sigma_1 \sigma_2 \\
    \rho \sigma_1 \sigma_2 & \sigma^2_2
  \end{bmatrix}$, with 
\begin{enumerate}
    \item Class 1: $\mu_1=25 \,\,\mu_2=50,\,\, \rho=0,\,\, \sigma_1=6,\,\, \sigma_2=3$
    \item Class 2: $\mu_1=38 \,\,\mu_2=40,\,\, \rho=0,\,\, \sigma_1=3,\,\, \sigma_2=3$
    \item Class 3: $\mu_1=45 \,\,\mu_2=35,\,\, \rho=0,\,\, \sigma_1=5,\,\, \sigma_2=5$
\end{enumerate}
for the problem with three ordered classes, and
\begin{enumerate}
    \item Class 1 $\mu_1=25 \,\,\mu_2=50,\,\, \rho=0\,\, \sigma_1=6,\,\, \sigma_2=3$
    \item Class 2 $\mu_1=30 \,\,\mu_2=45,\,\, \rho=0,\,\, \sigma_1=5,\,\, \sigma_2=5$
    \item Class 3 $\mu_1=38 \,\,\mu_2=40,\,\, \rho=0,\,\, \sigma_1=3,\,\, \sigma_2=3$
     \item Class 4 $\mu_1=45 \,\,\mu_2=35,\,\, \rho=0,\,\, \sigma_1=2,\,\, \sigma_2=3$
\end{enumerate}
for the problem with four ordered classes. 

\begin{figure}[ht] 
  \centering
  \includegraphics[width=0.6\textwidth]{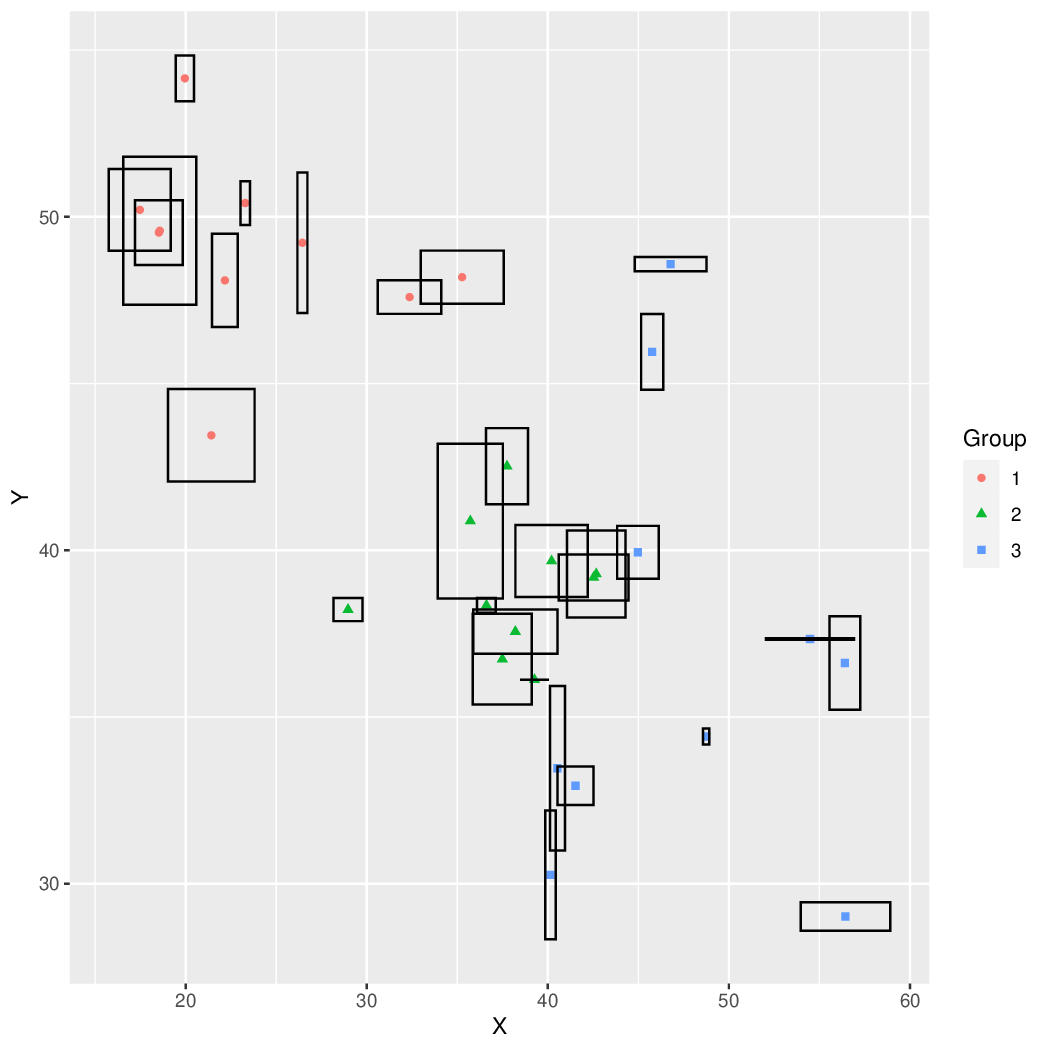}
  \caption{\textcolor{black}{Plot of 10 samples per group from the synthetic data with 3 classes. The rectangles denote the IVD,
and the dots are the corresponding midpoints.} \label{figsimu} }
\end{figure}

To build the interval-valued data set, each generated data point ($z_1, z_2$) is used as a seed of a vector of intervals (rectangle) defined as ($[z_1 - \gamma_1/2, z_1 + \gamma_1/2]$, $[z_2 - \gamma_2/2, z_2 + \gamma_2/2 ]$), where the parameters $\gamma_1$ and $\gamma_2$ are randomly drawn from a continuous uniform distribution on the interval $[1,5]$.  \textcolor{black}{Classes are balanced and several samples are shown in Fig. \ref{figsimu}.}

For both simulation designs, we generate 50 data sets, where 80\% of the samples are used for training and 20\% are used for testing, i.e. 240 (320) samples for training and 60 (80) samples for testing for the ordinal classification problem with 3 (4) classes. 

Table \ref{accsimu} contains a summary of the performance of each method for both simulation designs separately (second and third column) and jointly (last column). The maximum value in each column appears in bold. The p-values of the multiple comparison of means by Tukey contrasts are shown in Table \ref{Tukeysimu} for both simulation designs jointly. We include the following significance codes for the ranges of the p-values:   '***' means that the p-value is in the interval  [0, 0.001],     '**' means that the p-values are in the interval      (0.001, 0.01], '*' means that the p-value belongs to the interval     (0.01, 0.05], '.' means that the p-value is in the interval        (0.05, 0.1], while no significance code indicates that the p-value is in the interval    (0.1, 1]. Depending on the $\alpha$ level considered, we would say that there is a statistically significant difference between the mean accuracy values of both methods.

\begin{table}[ht]
\caption{\label{accsimu}Mean and standard deviation, in brackets, of accuracy (percentage) over 50 simulations for synthetic data.}
\centering
\begin{tabular}{cccc}
Method & 3 classes & 4 classes &  Global \\\hline
POLR & 88.2 (3.7) & 76.5 (4.5) & 82.4 (7.2)\\
OF& 88.1 (3.2) & {\bf 77.8} (4.5) & 83 (6.5) \\
LDA-ID & 88.2 (3.7) & 76.5 (4.5) & 82.4 (7.2) \\
FH+LDA-ID & 88.1 (3.2) & 76.1 (4.5) & 82.1 (7.2)\\
DI+ $wk$NN & 89.8 (3.5) & {\bf 77.8} (4.2) & 83.8 (7.2)\\
KPCA + POLR & 88.1 (3.2) &{\bf  77.8} (4.5) & 83 (6.5)\\
KIOF &  {\bf 90.7} (3.2) & 77.1 (4.3) & {\bf 83.9} (7.8)\\
POLR-I & 88.2 (3.7) & 76.5 (4.5) & 82.4 (7.2)\\
POLR-I2 & 88.1 (3.2) & {\bf 77.8} (4.5) & 83 (6.5)
\\\hline
\end{tabular}
\end{table}

\begin{table}[ht]
\caption{\label{Tukeysimu} P-values of Tukey simultaneous comparison  for synthetic data.}
\begin{tiny} 	
\centering
\setlength{\tabcolsep}{4pt}
\renewcommand{\arraystretch}{1.3}
\begin{tabular}{ccccccccc}
Method &   OF & LDA-ID &  FH+LDA-ID  & DI+ $wk$NN  & KPCA + POLR & KIOF & POLR-I  & POLR-I2 \\\hline
POLR &  .2235 & 1 &.5670 & .0021**& .2235& .0011**& 1& .2235 \\
OF& & .2235& .0737. & .0627.& 1& .0394*& .2235& 1 \\
LDA-ID &  & & .5670& .0021**& .2235& .0011**& 1& .2235 \\
FH+LDA-ID & & & & .0003***& .0737. & .0001***&.5670 & .0737. \\
DI+ $wk$NN & &&&& .0627.& .8415 & .0021** & .0627.\\
KPCA + POLR & &&&&& .0394* & .2235 & 1\\
KIOF & &&&&&& .0011** & .0394*\\
POLR-I & &&&&&&& .2235 
\\\hline
\end{tabular}
\end{tiny}
\end{table}

With 3 ordered classes, the best method is KIOF, with 90.7\%  mean accuracy, followed by DI+$wk$NN \textcolor{black}{(with $D^{EH}$)}, with 89.8\%  mean accuracy, and whose differences are statistically significant with respect to the rest of the methods. For the sake of brevity, we do not show the table of p-values for the simulation design with 3 classes, but the p-values of the Tukey contrasts for DI+ $wk$NN and KIOF with respect to the rest of the methods are below 3e-05. There is also a statistically significant difference between DI+ $wk$NN and KIOF (p-value = 0.0279). Therefore, two of our proposals show better performance than the na\"ive approaches. \textcolor{black}{Table \ref{statsimu3} shows more performance statistics for illustrative purpose, as our problems have balanced sample sizes and this kind of statistics are more critical in imbalanced situations. The
highest or second highest value in the majority (6) of columns are reached by DI+ $wk$NN and KIOF. For the F1-score columns, which combine precision and recall, the highest values correspond to FH+LDA-ID and DI+ $wk$NN for class 1 (KIOF is the third best performing method); KIOF and DI+ $wk$NN for class 2; and  KIOF and DI+ $wk$NN for class 3.}

\begin{table}[ht]
\caption{\label{statsimu3} \textcolor{black}{Mean of precision, recall and F1 for each class, over 50 simulations for synthetic data with three classes.  The maximum value in each column appears in bold.}}
\centering
\setlength{\tabcolsep}{5pt}
\begin{tabular}{cccccccccc}
Method & \multicolumn{3}{c}{Precision} & \multicolumn{3}{c}{Recall} &  \multicolumn{3}{c}{F1} \\
& 1& 2&3 &1 & 2&3 & 1 & 2&3\\
\hline
POLR & 98.2 & 80.0 & 86.8 & 96.8 & 85.6 & 82.2 & 97.5 & 82.3 & 84.0\\
OF& 94.0 & 82.0 & 88.5 &  99.6 & 84.5 & 79.9 & 96.6 & 82.6 & 83.4 \\
LDA-ID &  {\bf 100}  & 78.5 & 87.5 & 97.2 & 88.8& 79.3 & 98.6 & 83.0 & 82.8\\
FH+LDA-ID &  98.6 & {\bf 84.2} & 81.4 & {\bf 99.8} & 79.0 & {\bf 85.9} & {\bf 99.2} & 81.1 & 83.2\\
DI+ $wk$NN & 99.4 & 81.7 & 89.4 & 98.6 & {\bf 89.7} & 81.8 & 99.0 & 85.1 & 85.0 \\
KPCA + POLR & 97.6 & 80.1& 85.9 & 96.5 & 83.2 & 83.7 & 97.0 & 81.1 & 84.4 \\
KIOF & 99.1 & 82.8 & {\bf 90.9} & 98.0 & 90.7 & 83.8 & 98.5 & {\bf 86.2} & {\bf 86.9} \\
POLR-I &  98.4 & 80.2 & 86.0 & 96.9 & 84.3 & 83.3 & 97.6 & 81.8 & 84.1\\
POLR-I2& 98.2 & 79.7 & 86.9 & 96.8 & 85.8 & 81.8 & 97.5 & 82.3 & 83.9
\\\hline
\end{tabular}
\end{table}

With 4 ordered classes, four methods obtained the highest mean accuracy with 77.8\%  mean accuracy. 
These four methods are OF, DI+ $wk$NN, KPCA + POLR, and POLR-I2. The  method with the next highest mean accuracy is KIOF, whose mean accuracy is 77.1\%. As before, for the sake of brevity, we do not show the table of p-values for the simulation design with 4 classes, but the p-values of the Tukey contrasts for KIOF with respect to the four best methods are 0.0695. The p-values of the Tukey contrasts for the four best methods and all other methods except KIOF are below 0.001, so they are statistically significant. Therefore, three of our proposals together with the na\"ive method OF provide the best performance. 

When we consider both simulation designs jointly, the best performance is achieved by KIOF, with 83.9\%  mean accuracy. The second best  performance is achieved by DI+ $wk$NN, with 83.8\%  mean accuracy. The third best performance is achieved by  three methods, OF, KPCA + POLR, and POLR-I2, with 83\%  mean accuracy. The p-values of the Tukey contrasts for KIOF with respect to these three methods are 0.0394, while for all other methods (LDA-ID, FH + LDA- ID, POLR -I), they are below 0.0011, as can be seen in Table \ref{Tukeysimu}. Therefore, they are statistically significant. Analogously, the p-values of the Tukey contrasts for DI+ $wk$NN with respect to the three methods OF, KPCA + POLR, and POLR-I2 are 0.0627, while for all other methods (LDA-ID, FH + LDA-ID, POLR-I), they are below 0.0021. The method that
performs least well is FH+LDA-ID, with 82.1\%  mean accuracy, which is not statistically significantly different from POLR, LDA-ID, and POLR-I, with 82.4\%  mean accuracy.

In summary, our proposals KIOF and DI+$wk$NN are the methods that perform best, in fact, statistically significantly better than the rest of the methods for an $\alpha$ level of 0.1.

\subsection{Global Development data} \label{deve}

We consider development data from of 183 countries. The input features are two interval-valued variables based on the following two gender inequality indicators from \citet{wbdata}: a) Women Business and the Law Index Score (\href{https://data.worldbank.org/indicator/SG.LAW.INDX}{LAW}) (scale 1-100), which measures how laws and regulations affect women’s economic opportunities. Overall scores are calculated by taking the average score of each index (Mobility, Workplace, Pay, Marriage, Parenthood, Entrepreneurship, Assets, and Pension), with 100 representing the highest possible score; b) The percentage of seats held by women in national parliaments (\href{https://data.worldbank.org/indicator/SG.GEN.PARL.ZS}{GenPar}), which represents  the percentage of parliamentary seats in a single or lower chamber held by women.

The extremes of the interval-valued variables are the minimum and the maximum for both gender indicators between 2000 and 2021.

\begin{figure}[ht] 
  \centering
  \includegraphics[width=0.6\textwidth]{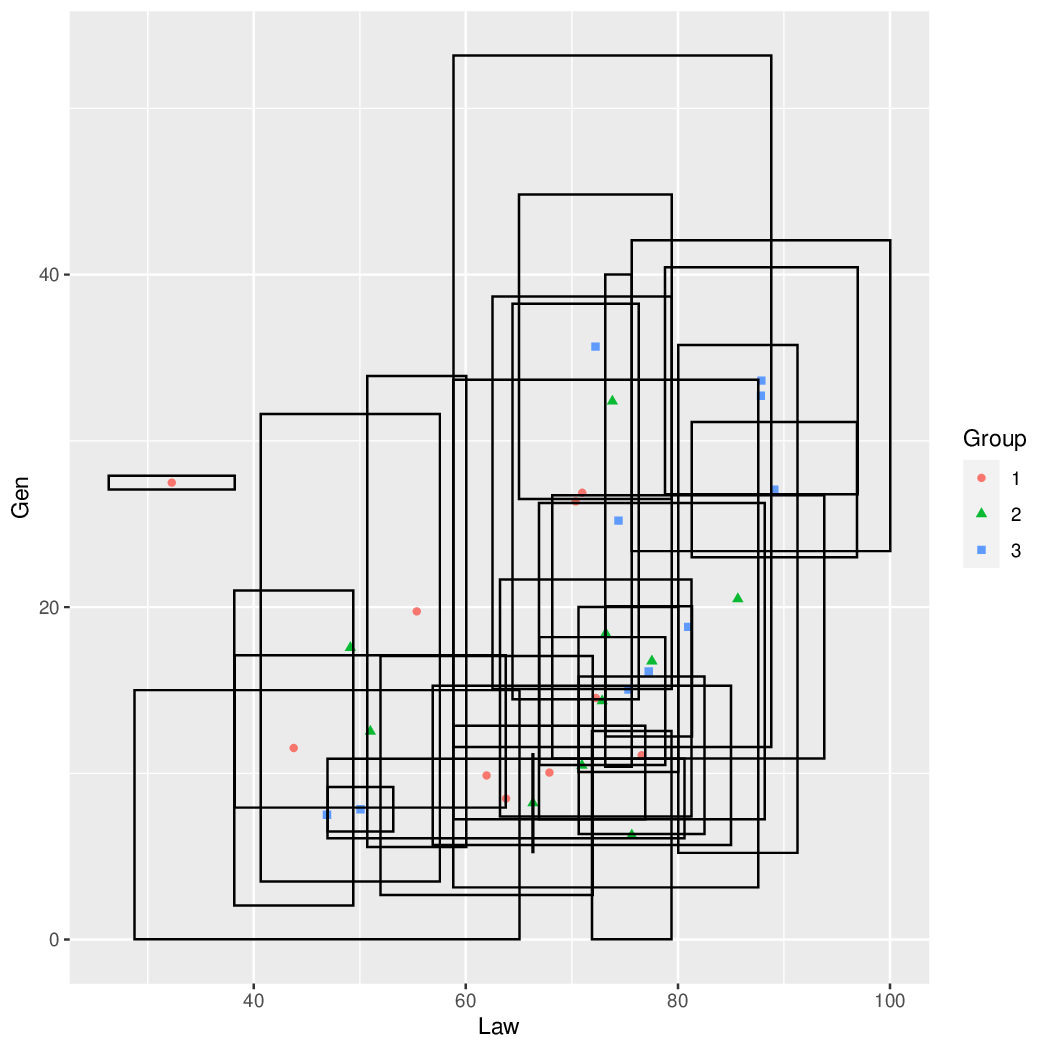}
  \caption{\textcolor{black}{Plot of 10 samples per group from the global development data with 3 classes. The rectangles denote the IVD,
and the dots are the corresponding midpoints.} \label{figpas} }
\end{figure}

As regards the output feature, the ordered factor is  built by dividing the Human-Development Index (HDI) for 2021 into certain percentiles  \citep{hdi}. The HDI is the geometric mean of three normalized indices: $\textrm{HDI}=\sqrt[{3}]{\textrm{LEI} \cdot \textrm{EI} \cdot \textrm{II}}$, where LEI stands for the Life Expectancy Index, EI for the Education Index, and II, the  Income Index.  HDI assesses having a long and healthy life, being knowledgeable, and having a decent standard of living. For the ordered categorical variable, we consider 5 possible levels from 3 to 7. For example, for the case of 3 ordered classes, the classes are defined by the labels $[0, L_{33}[,\, [L_{33}, L_{66}[,\, [L_{66}, L_{100}]$, where $L_i$ denotes the value of the $i$-th percentile of HDI. \textcolor{black}{Therefore, classes are balanced. Several samples are displayed in Fig. \ref{figpas}. In summary}, we have 5 data sets with the same inputs, but 5 different outputs, which have a different number of ordered categories ranging from 3 to 7. For each  of these 5 data sets, we use a Monte Carlo cross-validation, where 50 random splits of each data set are created. In each split, the data set is divided into training data (80\% of 183 countries) and validation data (20\% of 183 countries).

\begin{figure}[ht] 
  \centering
  \includegraphics[width=0.8\textwidth]{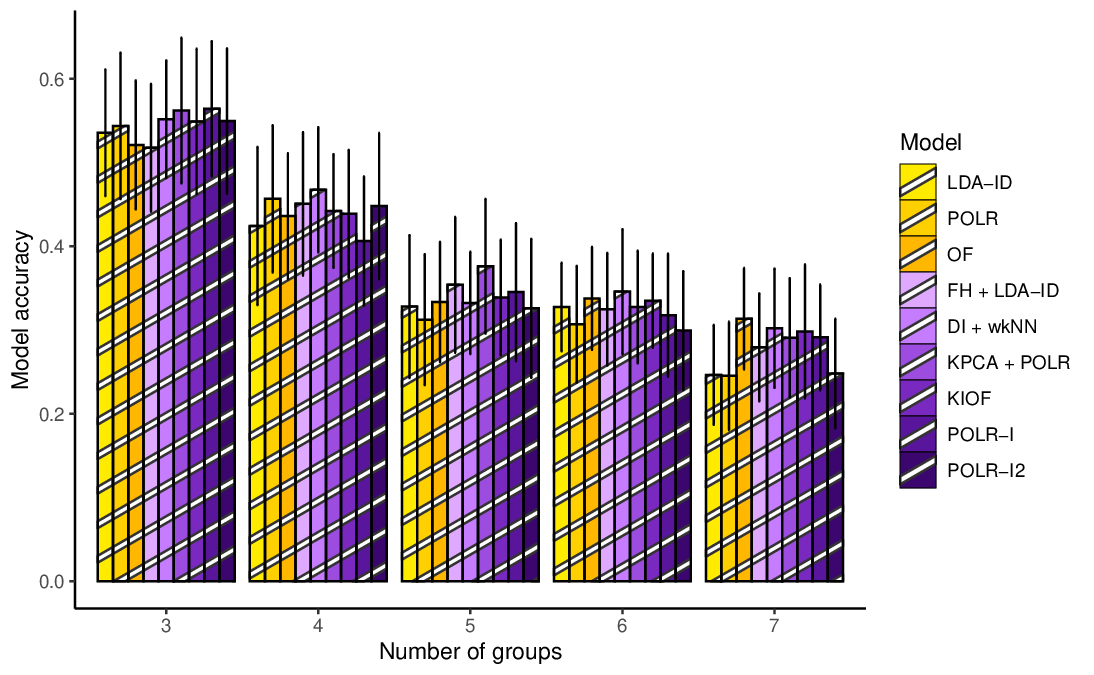}
  \caption{Mean and standard deviation of accuracy  over 50 simulations for each experimental design of the global development data. \label{countries} } 
\end{figure}

Table \ref{accshdi} shows a summary of the performance of each method for the experimental designs jointly, while Fig. \ref{countries} displays the performance
separately.  The p-values of the multiple comparison of means by Tukey contrasts are shown in Table \ref{Tukeyshdi} for the experimental designs jointly. As before, we include the significance codes. \textcolor{black}{Table \ref{statscountries3} shows performance statistics for three classes.}

\begin{table}[ht]
\caption{\label{accshdi} Mean and standard deviation, in brackets, of accuracy (percentage) over 50 splits of the 5 data sets from the global development data.}
\centering
\begin{tabular}{cc}
Method &  Global \\\hline
POLR & 37.31 (13.47)\\
OF&  38.84 (10.48)\\
LDA-ID & 37.24 (12.42)\\
FH+LDA-ID & 38.54 (11.46)\\
DI+ $wk$NN & 40.0  (11.77)\\
KPCA + POLR & 39.98 (12.14)\\
KIOF  & 39.21 (11.75)\\ POLR-I & 38.51 (12.31)\\
POLR-I2 & 37.43 (13.5)
\\\hline
\end{tabular}
\end{table}

\begin{table}[ht]
\caption{\label{Tukeyshdi} P-values of Tukey simultaneous comparisons for the global development data.}
\centering
\begin{tiny} 	
 \setlength{\tabcolsep}{3pt}
\renewcommand{\arraystretch}{1.3}
\begin{tabular}{ccccccccc}
Method  & OF & LDA-ID &  FH+LDA-ID  & DI+ $wk$NN  & KPCA + POLR & KIOF & POLR-I  & POLR-I2 \\\hline
POLR & .01734* & .91987 & .05605 .& 3.09e-05***& 3.58e-05*** & .00320** & .06283 . & .85367  \\
OF&  & .01315* & .63874 & .07291 . & .07843 .  & .56864 & .60324 & .02815*  \\
LDA-ID & & & .04432* & 1.98e-05*** & 2.30e-05*** & .00230** & .04990* & .77562 \\
FH+LDA-ID &  & & & .02369* & .02583* & .29861 & .95988 & .08428 .\\
DI+ $wk$NN &  & & & & .97325 & .22104 & .02076* & 6.80e-05*** \\
KPCA + POLR &  & & & & & .23395 & .02267* & 7.82e-05***\\
KIOF & & & & & & &  .27585 & .00571** \\
POLR-I & & & & & & & & .09371 .  
\\\hline
\end{tabular}
\end{tiny}
\end{table}

\begin{table}[ht]
\caption{\label{statscountries3} \textcolor{black}{Mean of accuracy (Acc.), precision, recall and F1 for each class, over 50 simulations  for the global development data with three classes.  The maximum value in each column appears in bold.}}
\centering
\color{black}
\setlength{\tabcolsep}{3.5pt}
\begin{tabular}{ccccccccccc}
Method & Acc & \multicolumn{3}{c}{Precision} & \multicolumn{3}{c}{Recall} &  \multicolumn{3}{c}{F1} \\
 & & 1& 2&3 &1 & 2&3 & 1 & 2&3\\
\hline
POLR & 54.4 & 56.8 & 45.1 & 61.1 & 61.0 & 38.2  & 66.6  & 58.1  & 39.7  & 62.0  \\
OF & 52.1 & 52.2 & 40.9 & 57.6 & {\bf 78.0} & 16.3 & 65.9 & 61.6 &23.7 & 60.2 \\
LDA-ID & 53.6 & 51.6 &49.9  &61.6  & 50.1 & 49.2 &64.5  &49.5  &48.1  &61.8 \\
FH+LDA-ID & 51.8 & 52.1 & 45.4 & 69.0 & 40.8 &{\bf 65.3}  &51.2  &44.2  & {\bf 52.8} & 57.6\\
DI+ $wk$NN & 55.2 &  57.9& 44.5 & {\bf 71.3} & 58.3 &54.6  &53.9  &56.7  & 48.0 &60.5 \\
KPCA + POLR & 56.2 & 54.2 & 48.7 & 69.9 & 72.8 & 36.6 & 62.2 & 61.0 & 39.8 & {\bf 65.0} \\
KIOF &  54.9& 55.7 & 45.9 & 63.6 & 76.5 & 30.3 & 61.6 & {\bf 63.5} & 35.7 & 61.3 \\
POLR-I & {\bf 56.4} & {\bf 59.2} & {\bf 50.3}  & 62.9  & 67.8 & 41.8 & 63.8 & 62.4 &  44.0 & 61.5\\
POLR-I2 & 55.0 & 56.6 & 45.9 & 63.8  & 62.2 & 37.9 & {\bf 68.8} & 58.2 & 39.2 & 64.5 \\\hline
\end{tabular}
\end{table}

According to the results in Table \ref{accshdi}, the best methods are DI+ $wk$NN \textcolor{black}{(with $D^{EH}$)} and KPCA + POLR, with 40\% and 39.98\%   mean accuracy, respectively. The third best performance is achieved by KIOF, with 39.21\%  mean accuracy. They are not statistically significantly different according to the p-values of the Tukey contrasts in Table \ref{Tukeyshdi}. However, DI+ $wk$NN and KPCA + POLR are statistically significantly different with respect to the rest of the methods. OF is the fourth best  method in terms of performance, with 38.84\%  mean accuracy. This  na\"ive method achieves results similar to FH+LDA-ID and POLR-I, with 38.54\% and 38.51\%  mean accuracy. The  methods that perform worst are LDA-ID, POLR, and POLR-I2, with 37.24\%, 37.31\%,  and 37.43\%  mean accuracy, respectively. These three methods form a homogeneous group that is, statistically significantly different from the rest of the methods at level 0.1. 

According to the results of Fig. \ref{countries}, the best accuracy with 3 ordered classes is provided by POLR-I, with 56.4\% mean accuracy. For the sake of brevity, we do not show the table of p-values for each experimental design, but in this case, no statistically significant difference is found between POLR-I and 
  DI+ $wk$NN, KPCA + POLR, KIOF, and POLR-I2 at level 0.1, i.e. we find a statistically significant difference between POLR-I and all the na\"ive approaches (LDA-ID, POLR, OF), together with FH+LDA-ID. \textcolor{black}{If F1-score rankings in Table \ref{statscountries3} are considered, the methods with jointly lowest rankings (best performing methods) in the three classes coincide with those methods with the highest accuracies (POLR-I and KPCA + POLR).}

  \textcolor{black}{Following the results of Fig. \ref{countries},} the best accuracy with 4 ordered classes is provided by DI+ $wk$NN, with 46.8\%  mean accuracy. DI+ $wk$NN is statistically significantly different from LDA-ID, OF, KPCA + POLR, KIOF, and POLR-I at level 0.1. The best accuracy with 5 ordered classes is provided by KPCA + POLR, with 37.6\%  mean accuracy. KPCA + POLR is statistically significantly different from the rest of the methods at level 0.1. The best accuracy with 6 ordered classes is provided by DI+ $wk$NN, with 34.6\%  mean accuracy. DI+ $wk$NN is statistically significantly different from  POLR, FH+LDA-ID, POLR-I, and POLR-I2 at level 0.1. The best accuracy with 7 ordered classes is provided by OF, with 31.4\%  mean accuracy. OF is statistically significant different from LDA-ID, POLR, FH+LDA-ID, KPCA + POLR, POLR-I, and POLR-I2 at level 0.1. In summary, except for one simulation design, DI+ $wk$NN is among the best methods in all situations.


\subsection{Catalan meteorological data} \label{meteo}
\begin{figure}[ht] 
  \centering
  \includegraphics[width=0.8\textwidth]{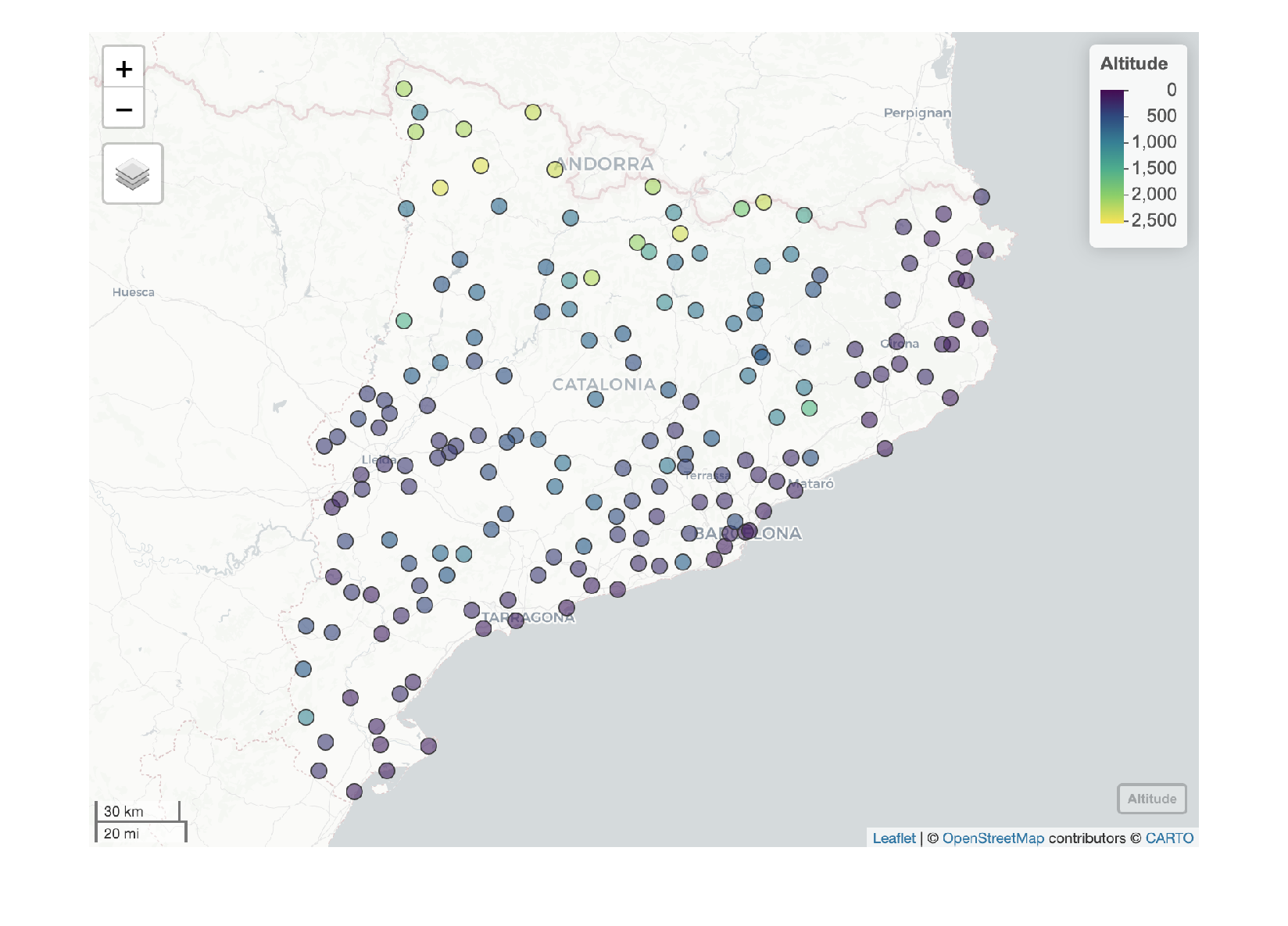}
  \caption{Map of Catalonia (Spain) with the situation of the Catalan weather stations and their altitude. \label{stationsfig} } 
\end{figure}

\begin{figure}[ht] 
  \centering
  \begin{tabular}{cc}
     \includegraphics[width=0.5\textwidth]{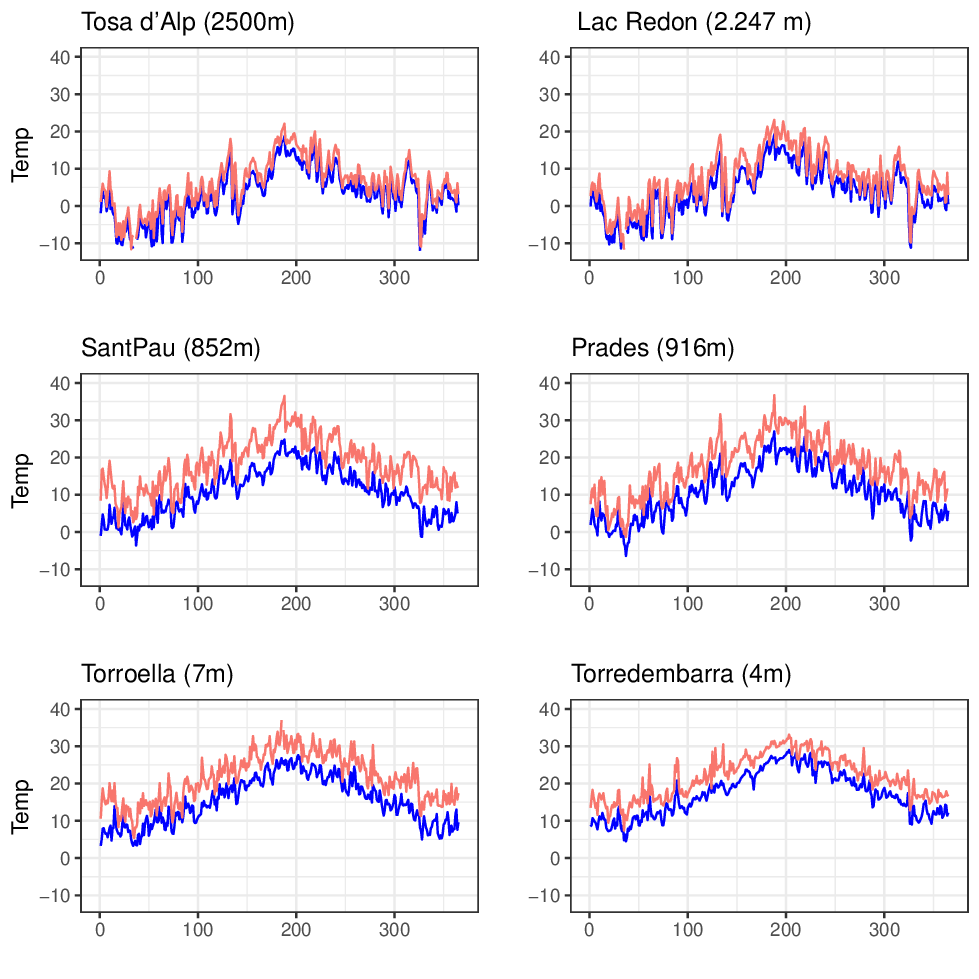}  & \includegraphics[width=0.5\textwidth]{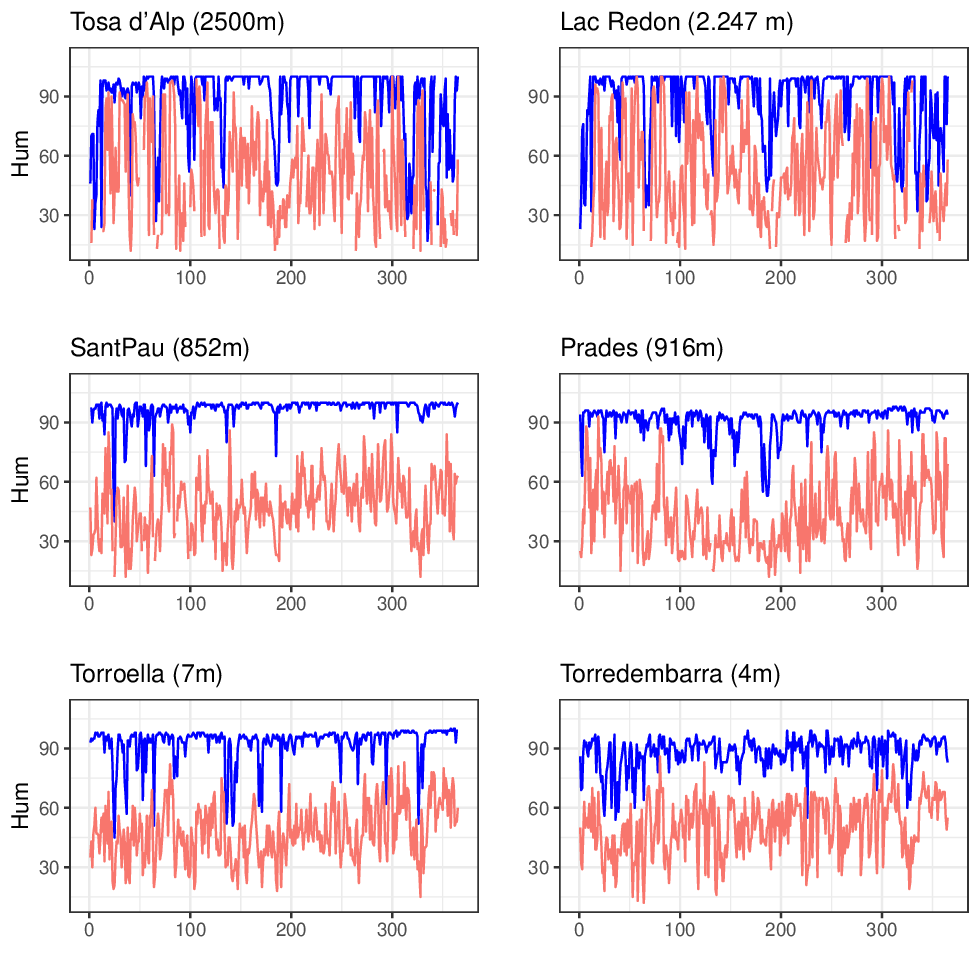} \\
      a) & b)
  \end{tabular}
    \caption{\textcolor{black}{a) Minimum (blue) and and maximum (red) daily temperatures on the Celsius scale and b) minimum (blue) and maximum (red) daily relative humidity values (right) of a sample of six stations. Weather stations of high, medium and low altitude appear in top, middle and bottom rows, respectively.} \label{funmet} } 
\end{figure}

We consider data from 160 Catalan weather stations in 2015 (see Fig. \ref{stationsfig}). These data were provided by the Servei Meteorològic de Catalunya (Catalan Meteorological Service). The input features are two functional interval-valued variables observed at 365 points: a) the  minimum  and maximum daily temperatures measured in degrees Celsius for each of the 365 days; and b) the minimum and maximum daily relative humidity values measured in percentage for each of the 365 days. \textcolor{black}{Fig. \ref{funmet} displays a sample of these functions.} As both variables are measured in non-compatible units, each functional variable is standardized, so that both variables have an equal weight in the methods that use distances. We consider the average daily temperatures of each day and weather station, which were also provided by the Servei Meteorològic de Catalunya. Then, functional means and variances are defined daily across weather stations \citep{Ramsay05}. The same procedure is carried out for relative humidity. The functional averages are subtracted from the respective functions, then the functions are
divided by the respective standard deviation functions.

 Note that our two interval-valued functional variables are equivalent to 730 (365 $\times$ 2) interval-valued variables, which are also highly correlated between neighboring days. This means that some methods fail and cannot be used with these data. In particular, the methods that fail are LDA-ID, POLR, OF, FH+LDA-ID, POLR-I, and POLR-I2, i.e. all the methods except DI+ $wk$NN \textcolor{black}{(with $D^{FEH}$ for bivariate IVF)}, KPCA + POLR, and KIOF. We have tried to solve this problem by sampling the days, and rather than using 365 days, considering only one in every 30 days  for the methods that fail. Therefore, we work with two interval-valued functional variables observed on  12 days, rather than 365 days, for methods LDA-ID, OF, and FH+LDA-ID, which is equivalent to 24 (12 $\times$ 2)  interval-valued variables. However, methods POLR, POLR-I, and POLR-I2 continue to fail, and, therefore, they are not considered in this problem. In summary, for this problem, we work with the full data set with DI+ $wk$NN, KPCA + POLR, and KIOF, and with a time sampled data set for LDA-ID, OF, and FH+LDA-ID.

As regards the output feature, the ordered factor is  built from the division into certain percentiles of altitude of each weather station. As before, for the ordered categorical variable, we consider 5 possible levels from 3 to 7. For example, for the case of 3 ordered classes, the \textcolor{black}{balanced} classes are defined by the labels $[0, L_{33}[,\, [L_{33}, L_{66}[,\, [L_{66}, L_{100}]$, where $L_i$ denotes the value of the $i$-th percentile of altitude. Therefore, we have 5 data sets with the same inputs, but 5 different outputs, which have a different number of ordered categories ranging from 3 to 7. For each of these 5 data sets, we use a Monte Carlo cross-validation, where 50 random splits of each data set are created. In each split, the data set is divided into training data (80\% of the weather stations) and validation data (20\% of the weather stations).

\begin{figure}[ht] 
  \centering
  \includegraphics[width=0.8\textwidth]{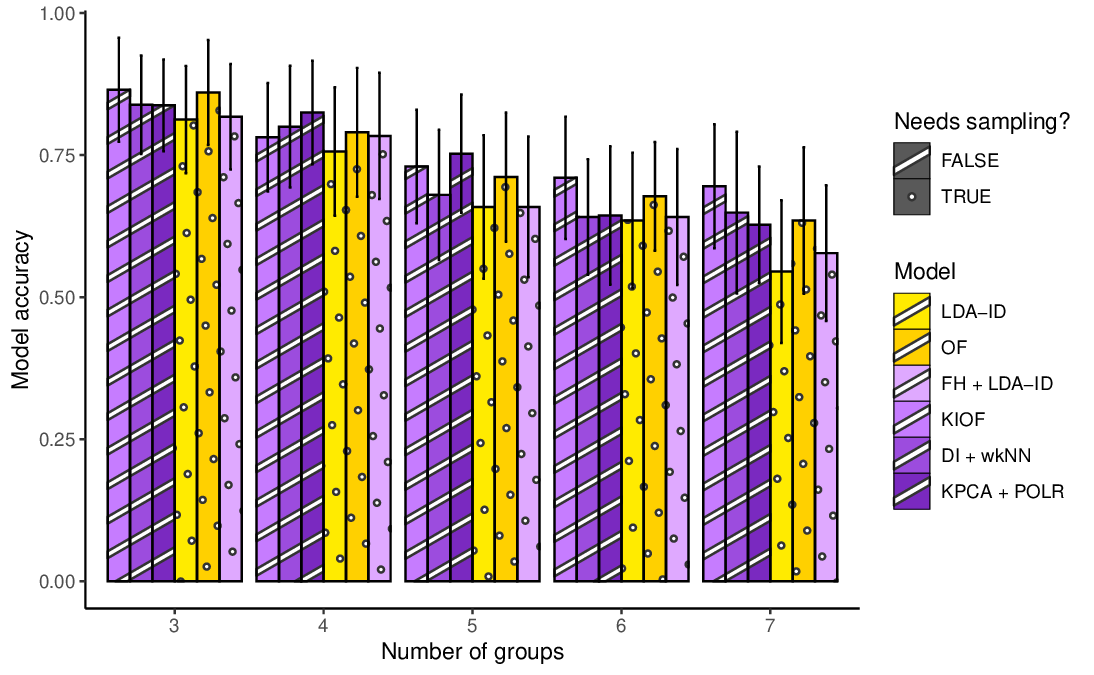}
  \caption{Mean and standard deviation of accuracy  over 50 simulations for each experimental design of the  meteorological data. \label{meteofig} } 
\end{figure}

Table \ref{accmet} shows a summary of the performance of each method for the experimental designs jointly, while Fig. \ref{meteofig} displays the performance
separately.  The p-values of the multiple comparison of means by Tukey contrasts are shown in Table \ref{Tukeymet} for the experimental designs jointly. As before, we include the significance codes. \textcolor{black}{Table \ref{statsmeteo3} shows performance statistics for three classes.}

\begin{table}[ht]
\caption{\label{accmet} Mean and standard deviation, in brackets, of accuracy (percentage) over 50 splits of the 5 data sets from the meteorological data.}
\centering
\begin{tabular}{cc}
Method &  Global \\\hline
OF& 73.48 (13.55)\\
LDA-ID & 68.15 (14.89) \\
FH+LDA-ID & 69.58 (14.48) \\
DI+ $wk$NN & 72.18 (13.8)\\
KPCA + POLR & 73.72 (13.35)\\
KIOF & 75.62 (11.77)\\\hline
\end{tabular}
\end{table}

\begin{table}[ht]
\caption{\label{Tukeymet} P-values of Tukey simultaneous comparisons  for the meteorological data.}
\begin{footnotesize}
\centering
\renewcommand{\arraystretch}{1.2}
\setlength{\tabcolsep}{4pt}
\begin{tabular}{cccccc}
Method   & LDA-ID &  FH+LDA-ID  & DI+ $wk$NN  & KPCA + POLR & KIOF \\\hline
OF& 2.98e-08*** & 4.72e-05*** &  .17387 & .79363 & .02459* \\
LDA-ID &  &  .13608 & 2.68e-05*** & 6.65e-09*** & 9.77e-15***\\
FH+LDA-ID & & & .00659** & 1.50e-05*** & 3.23e-10*** \\
DI+ $wk$NN & & & &  .10498 &  .00032***\\
KPCA + POLR & & & & & .046947*
\\\hline
\end{tabular}
\end{footnotesize}
\end{table}

\begin{table}[ht]
\caption{\label{statsmeteo3} \textcolor{black}{Mean of accuracy (Acc.), precision, recall and F1 for each class, over 50 simulations  for the meteorological data with three classes.  The maximum value in each column appears in bold.}}
\centering
\color{black}
\setlength{\tabcolsep}{3.5pt}
\begin{tabular}{ccccccccccc}
Method & Acc & \multicolumn{3}{c}{Precision} & \multicolumn{3}{c}{Recall} &  \multicolumn{3}{c}{F1} \\
 & & 1& 2&3 &1 & 2&3 & 1 & 2&3\\
\hline
OF & 86.0  & 82.0 & {\bf 80.5} & 92.2 & {\bf 95.9} &71.1  & {\bf 88.8} & 87.5 & 74.5& {\bf 89.8} \\
LDA-ID &  81.3& 82.5 & 71.3 & 93.4 & 85.5  & 75.8 & 83.2 & 82.6 &72.1 & 86.8  \\
FH+LDA-ID & 81.8 & 83.9 & 71.8 & 93.5 & 88.4 & 76.2 & 82.4 & 84.6 &72.6 &86.5  \\
DI+ $wk$NN & 83.9 & 89.2 & 71.6 & {\bf 97.0} & 85.5 & {\bf 86.7} & 81.6 & 86.2 & 78.2 & 87.6 \\
KPCA + POLR &83.8  & 88.0 & 72.3 & 90.1 & 89.2 & 77.4 & 84.6 & 87.6 &75.5 & 85.8  \\
KIOF & {\bf 86.5} &  {\bf 90.5}& 77.8 & 93.4 & 93.5 & 84.9  & 82.8 & {\bf 91.1} & {\bf 80.1} & 86.1 \\\hline
\end{tabular}
\end{table}

According to the results in Table \ref{accmet}, the best method is KIOF, with 75.62\%  mean accuracy. KIOF is statistically significantly different from the rest of the methods according to the p-values of the Tukey contrasts in Table \ref{Tukeymet}. The following best  methods in terms of performance are KPCA + POLR, OF, and DI+ $wk$NN, with 73.72\%, 73.48\%, and 72.18\%  mean accuracy, respectively. These three methods form a homogeneous group that is statistically significantly different from the rest of the methods. The  methods that perform the worst are LDA-ID and FH+LDA-ID, with 68.15\% and 69.58\%  mean accuracy, respectively. These two methods form another homogeneous group  that is statistically significantly different from the rest of the methods. 

According to the results in Fig. \ref{meteofig}, the best accuracy with 3 ordered classes is provided by KIOF, with 86.5\%  mean accuracy. For the sake of brevity, we do not show the table of p-values for each experimental design, but in this case, no statistically significant difference is found between KIOF and OF at level 0.05. \textcolor{black}{The highest or second highest value in the
majority (6) of columns (not including accuracy) in  Table \ref{statsmeteo3} is reached by KIOF. }

\textcolor{black}{Following the results of Fig. \ref{meteofig}, } the best accuracy with 4 ordered classes is provided by KPCA + POLR, with 82.5\%  mean accuracy. No statistically significant difference is found between  KPCA + POLR and DI+ $wk$NN at level 0.05. The best accuracy with 5 ordered classes is provided again by KPCA + POLR, with 75.2\%  mean accuracy. No statistically significant difference is found between  KPCA + POLR and KIOF at level 0.05. The best accuracy with 6 ordered classes is provided by KIOF, with 71\%  mean accuracy. No statistically significant difference is found between KIOF and OF at level 0.05.  The best accuracy with 7 ordered classes is provided again by KIOF, with 69.5\%  mean accuracy. KIOF is statistically significantly different from the rest of the methods at level 0.05. In summary, except for one simulation design, KIOF is among the best methods in all  situations. 

\subsection{Discussion} Let us consider  the results of the simulated and real data sets together. KIOF and DI + $wk$NN are the best methods (Table \ref{accsimu}) for simulated data. DI + $wk$NN, KPCA + POLR, and KIOF are the best methods (Table \ref{accshdi}) for the global development data, while KIOF is the best method for the meteorological data. Therefore, KIOF appears to be the best methodology in all three situations. KIOF is a nonparametric and highly nonlinear method. Appart from KIOF, DI + $wk$NN and KPCA + POLR also seem better alternatives than FH + LDA-ID, POLR-I, and POLR-I2. These three methods are parametric, unlike KIOF and DI + $wk$NN, which are nonparametric methods. KPCA + POLR combines a nonparametric part with a parametric method. 

Taking order and interval-valued information into account is beneficial. Although OF is quite competitive despite being a na\"ive approach, KIOF that combines OF and the use of the interval-valued information improves the performance. The other two  na\"ive methods, POLR and LDA-ID do not perform so well, being among the worst methods in all the data sets. FH + LDA-ID is only statistically significantly better than LDA-ID for the global development data. Note that it depends on the class probabilities returned by LDA-ID.

For the functional case, it is clear that KIOF, DI + $wk$NN, and KPCA + POLR are the best options, since for the other methods we have to discard some information by sampling.

\section{Conclusions \label{conclusiones}}
We have proposed six methodologies (FH + LDA-ID, DI + $wk$NN, KPCA + POLR, KIOF, POLR-I, and POLR-I2) for computing ordinal classification in two different cases, with interval-valued data and functional interval-valued data as inputs. To the best of our knowledge, this is the first time these issues have been addressed. We have made an extensive comparative study with simulated and real data sets and different experimental setups regarding the number of levels of output. 

Although there is no single method that always performs  the best in all possible data sets, there are some methods that are more recommendable. KIOF has returned excellent results  with both interval-valued data and functional interval-valued data. DI + $wk$NN and KPCA + POLR are also recommendable in both cases. 

As future work, more methods could be explored. For example, we use KPCA + POLR, but another ordinal classification method could be used after KPCA rather than POLR. POLR with variable selection could also be  used. Another line of future work would be to work with interval-valued mixed data, with function and vector parts, or to extend the work to other symbolic data. However, applications are also one of the main directions for future work \textcolor{black}{since many times interval-valued information is not exploited \citep{10161717}}. \textcolor{black}{Other ways to explore are dealing with incomplete interval-valued data \citep{qi2021reliable} and imbalanced interval-valued data \citep{qi2023agurf}. Note that we obtain ordinal classification problems by discretizing the response into $Q$ different classes with equal frequency. We prefer to assess the performance in this more controlled environment since this is the first time that ordinal classification is addressed with IVD and IVF. }



\section*{CRediT authorship contribution statement} 
A. A.: Data curation, Formal analysis, Investigation, Software, Visualization, Writing - review \& editing. M. M-G.: Data curation, Formal analysis, Investigation, Software, Visualization, Writing - review \& editing.   I.E.: Conceptualization, Funding acquisition, Methodology, Project administration,  Supervision, Writing - original draft, Writing - review \& editing.

\section*{Declaration of competing interest}
The authors declare that they have no known competing financial interests or personal relationships that could have  influenced the work reported in this paper.

\section*{Acknowledgments}
The authors would like to thank the Servei Meteorològic de Catalunya for providing them with the meteorological data.

This research was partially supported by the Spanish Ministry of Universities (FPU grant FPU20/0182), Spanish Ministry of Science and Innovation (PID2022-141699NB-I00, PID2020-118763GA-I00 and PID2020-118071GB-I00), CIGE/2022/066 from Generalitat Valenciana and UJI-B2020-22 and  UJI-A2022-12 from Universitat Jaume I, Spain.


\end{document}